\documentclass[a4paper]{jpconf}
\usepackage{graphicx}
\usepackage{amssymb}

\def\a{\alpha}
\def\b{\beta}

\def\d{\delta}

\def\g{\gamma}

\def\k{\kappa}
\def\l{\lambda}
\def\m{\mu}

\def\s{\sigma}
\def\t{\tau}

\def\D{\Delta}

\def\G{\Gamma}

\def\L{\Lambda}
\def\O{\Omega}

\def\ve{\varepsilon}

\def\dg{\dagger}                                     

\def\be{\begin{equation}}
\def\ee{\end{equation}}

\def\bea{\begin{eqnarray}}
\def\eea{\end{eqnarray}}

\begin{document}
\title{Beyond the Standard Model with leptogenesis and neutrino data}

\author{Pasquale Di Bari}

\address{Department of Physics and Astronomy, Highfield, 
University of Southampton, Southampton SO17 1BJ, UK}

\ead{P.Di-Bari@soton.ac.uk}

\begin{abstract}
In this short review \footnote{Compendium of talks given at Neutrino 2016 and NuFact 2016.} I discuss
how high energy (type I) seesaw models can be nicely embedded within grand-unified models and reproduce the observed matter-antimatter asymmetry with leptogenesis.
In particular, after discussing general features and results in leptogenesis,
 I focus on $SO(10)$-inspired leptogenesis and on a particular solution,
the strong thermal $SO(10)$-inspired solution, that provides an interesting way to
understand neutrino mixing parameters: the non-vanishing 
reactor mixing angle, the emerging negative sign of $\sin\delta$ and 
the slight hints favouring normally ordered neutrino masses
and an atmospheric mixing angle in the first octant. I also briefly discuss
leptogenesis within two right-handed seesaw neutrino models. In this case a
a third decoupled right-handed neutrino can
provide a candidate for very heavy cold decaying dark matter
produced from right-handed neutrino mixing with a mass in the TeV-EeV range
and its decays would give a contribution to the IceCube high energy neutrino events 
in addition to an astrophysical component. 
\end{abstract}

\section{Introduction}

The $\Lambda$CDM cosmological model has so far resisted all experimental efforts 
to unearth sound evidences motivating a modification or an extension.
Therefore, it nowadays provides a minimal model able to describe all cosmological observations 
\cite{WMAP,Planck}.  However, within the Standard Model of particle physics
and fundamental interactions (SM), we cannot explain the nature and origin of some
of the features of the $\Lambda$CDM, in particular the observed matter-antimatter asymmetry 
of the Universe and the necessity of a non-baryonic dark matter component. 
For this reason these {\em cosmological puzzles} have to be regarded as
strong motivations for new physics beyond the SM. Neutrino masses and mixing 
also call for an extension of the SM and it is then reasonable that the same 
new physics should also address the cosmological puzzles.

Here I focus on a simple extension based on the introduction of right-handed (RH) neutrinos
with Yukawa couplings, generating a Dirac mass term for neutrinos as for the other massive fermions,
and an additional Majorana mass term.  This simple extension leads to the seesaw
formula for the light neutrino masses and mixing and opens the opportunity to solve some of the 
cosmological puzzles in quite a minimal way that moreover can be easily embedded within well
motivated realistic models beyond the SM such as grand-unified models.

\subsection{Baryon asymmetry of the Universe}

The small CMB  deviations from a thermal equilibrium distribution 
combined with stringent constraints from cosmic rays place a tight
upper bound on the abundance of primordial ordinary anti-matter in our observable
Universe \cite{glashow}. The energy density of 
ordinary matter is today dominated  by baryons (in the form of nucleons)
and, therefore, the baryon contribution to the energy density of the Universe can be regarded
as a measure of the matter-antimatter asymmetry of the Universe. 

This is one of the cosmological
parameters that is more precisely and accurately measured by CMB temperature anisotropies, since
 it directly enters an expression of the sound velocity in the primordial plasma at recombination, affecting 
 quite remarkably the height of the CMB acoustic peaks \cite{hudodelson}. The {\em Planck} satellite
 collaboration finds for the baryon abundance \cite{Planck}
 \be
 \Omega^{(CMB)}_{B0} h^2 = 0.02222 \pm 0.00023 \,  .
 \ee
 This can be translated into a measurement of the {\em baryon-to-photon number ratio} 
 \be
 \eta_{B0} \equiv {n_{B0} \over n_{\g 0}} \simeq {\Omega_{B 0}\,\varepsilon_{\rm c 0} \over 
 m_{N}\,n_{\gamma 0}} \simeq 273.5 \times 10^{-10}\, \Omega_{B 0}\,h^2 
\Rightarrow \eta_{B 0}^{(CMB)} =(6.05 \pm 0.06) \times 10^{-10} \,  ,
 \ee
 where $m_N$ is the (properly averaged) mass of nucleons, $\ve_{\rm c}$
 is the critical energy density of the Universe and, as usual, the subscript ``$0$" indicates
 quantities at present. 
 
\subsection{Neutrino masses and mixing}

Neutrino mixing experiments are well explained, barring anomalies hinting at possible light
sterile neutrino states, by neutrino mixing among three active neutrino mass eigenstates with masses
$m_1 < m_2 < m_3$ and with mass squared differences given in the case of  normal (inverted) ordering  by
$m^2_3 - m^2_1 \equiv m_{\rm atm}^2 \simeq (0.05 \, {\rm eV})^2$  
and $m^2_2 - m_1^2 \, (m^2_3 - m^2_2) \equiv m^2_{\rm sol} \simeq (0.009\,{\rm eV})^2$ 
\cite{talkmarrone}.
Neutrino flavour eigenstates can be in general expressed as an admixture of neutrino mass eigenstates 
described by a leptonic mixing matrix $U$ such that $\nu_{\a} = \sum_i U_{\a i}\, \nu_i$ \cite{PMNS}.

Latest neutrino oscillation experiments global analyses find for the mixing angles and the 
Dirac phase $\d$, in the case of NO,
the following best fit values and $1\s$ errors ($3\s$ ranges) \cite{talkmarrone}: 
\bea\label{expranges}
\theta_{13} & = &  8.4^{\circ}\pm 0.2^{\circ} \, \;\;  (7.8^{\circ}\mbox{---}\, 9.0^{\circ}) \,  , \\ \nonumber
\theta_{12} & = &  33^{\circ}\pm 1^{\circ} \,  \;\;  (30^{\circ}\mbox{---}\, 36^{\circ})  \,  , \\ \nonumber
\theta_{23} & = &  {41^{\circ}} \pm {1^{\circ}} \,  \;\;  (38^{\circ}\mbox{---}\, 51^{\circ})  \,  ,  \\ \nonumber
\d & = &  {-108^{\circ}} \pm {36^{\circ}} \,  \;\;  (-207^{\circ}\mbox{---}\,27^{\circ})  \, .
\eea 
It is interesting that there is already a $3\s$ exclusion interval $\d\ni [27^\circ,153^{\circ}]$
and that  $\sin\d > 0$ is excluded at more than $2\s$ clearly favouring  $\sin\d < 0$
(a lower statistical significance is found in \cite{nufit}).  There are no 
experimental constraints on the Majorana phases and there is no signal from $00\nu\b$ experiments
that, therefore, place an upper bound on the $00\nu\b$ effective neutrino
mass $m_{ee}$. Currently, the most stringent reported upper bound comes from the KamLAND-Zen collaboration 
finding, at $90\%\,{\rm C.L.}$,  $m_{ee} \leq (61 \mbox{--} 165)\,{\rm meV}$  \cite{kamlandzen},
where the range accounts for nuclear matrix element uncertainties.

Cosmological observations place an upper bound on the 
sum of the neutrino masses, in particular the {\em Planck} satellite collaboration 
finds 
\footnote{Recently the {\em Planck} collaboration has presented new results 
\cite{planck16} based on a  reduction of
large-scale effects in high frequency polarisation maps  obtaining for the upper bound on neutrino
masses, combining {\em Planck} data with BAO information, $\sum m_i \lesssim 0.17\,{\rm eV}$ 
corresponding for NO (IO) to an upper bound $m_1 \lesssim  50\,(42)\,{\rm meV}$.}
$\sum_i m_i \lesssim 230\,{\rm meV}$ at $95\% {\rm C.L.}$
\cite{Planck} that, taking into account 
the measured values of the  solar and atmospheric neutrino mass scales, translates into an upper bound on the 
lightest neutrino mass $m_1 \lesssim 70\,{\rm meV}$. 

\subsection{Minimally extended SM}

In order to account for neutrino masses and mixing,
one could extend the SM minimally, just adding RH neutrinos with Yukawa couplings 
$h$ giving an additional term $-{\cal L}^{\nu}_{Y}= \bar{\nu}_L\,h \, \nu_R \, \phi$, as for the other 
fermions. After spontaneous symmetry breaking this 
generates a neutrino Dirac mass term $-{\cal L}^{\nu}_{\rm mass}= \bar{\nu}_L\, m_D \, \nu_R$, where 
$m_D = v \,  h$ is the neutrino Dirac mass matrix. It is always possible to find
two unitary transformations $V_L$ and $U_R$ that, acting respectively on LH and RH neutrino fields,
bring to the Yukawa basis, where the neutrino Dirac mass matrix is diagonal and given by
$D_{m_D} \equiv {\rm diag}(m_{D1},m_{D2},m_{D3})$, with $m_{D1} \leq m_{D2} \leq m_{D3}$,
in  a way that one can write (singular value decomposition)
\be
m_D = V^{\dg}_L \, D_{m_D} \, U_R \,  .
\ee
Within this picture,  the neutrino masses would be simply given by
the eigenvalues of the neutrino Dirac mass matrix, 
$m_{\nu i}= m_{Di}$ and the leptonic mixing matrix by $U=V^{\dagger}_L$.
However, this minimal extension does not address different issues
\footnote{This of course does not exclude the possibility that neutrino are of Dirac nature,
but this minimal picture has to be supplemented by further ingredients such as extra-dimensions
within a Randall-Sundrum setup \cite{neubert}.}: 
\begin{itemize}
\item[-] why neutrino masses are much lighter than all other massive fermions; 
\item[-] why we observe much large mixing angles  in $U$ compared to the quark sector;
\item[-] the cosmological puzzles;
\item[-]  why there is not a Majorana mass term in addition to the Dirac mass term.
\end{itemize}

\subsection{Seesaw mechanism}
 
Neutrinos only carry lepton number as a global charge and, therefore, having introduced
RH neutrinos and without modifying the SM Higgs sector, one can also have, in addition to the
Dirac mass term, a right-right  Majorana mass term without conflicting with any experimental bound. 
This term would violate lepton number at tree level and can have interesting phenomenological 
 consequences potentially testable. In this way after spontaneous symmetry breaking the total neutrino mass
 term now would read
 \be
 -{\cal L}^{\nu}_{\rm mass} = 
 (\bar{\nu}^c_L, \bar{\nu}_R) \,
 \left(
 \begin{array}{cc}
 0 & m_D \\
m_D^T & M
\end{array}
 \right) \,
 \left(
 \begin{array}{c}
\nu_L   \\
\nu^c_R
\end{array}
 \right) + \mbox{\rm h.c.}   \;\;\;   .
 \ee
 In the {\em seesaw limit}, $M \gg m_D$, the mass spectrum splits  into 2 sets: a set of three 
 light neutrinos (dominantly LH) with masses given by the seesaw formula 
 \be
 {\rm diag}(m_1,m_2,m_3) = U^{\dg}\,m_D\,M^{-1}\,m_D^T\,U^{\star} \,  ,
 \ee
 and a set of $N$ very heavy (dominantly RH) neutrinos with masses $M_1 \leq \dots \leq M_N$
(almost) coinciding with the eigenvalues of the Majorana mass matrix $M$. The number of RH neutrinos
$N$ cannot be lower than two in order to reproduce the solar and the atmospheric neutrino mass scale
but, model independently, there is no upper bound. However, for definiteness and since we will be interested
in $SO(10)$-inspired models, we will refer to the case $N=3$. 

These new very heavy RH neutrinos can now play a cosmological role, as we will see, being either 
responsible for the matter-antimatter asymmetry or providing a candidate for
cold dark matter or both. 

\section{Minimal scenario of leptogenesis}

The minimal scenario of leptogenesis \cite{fy}  relies on two main assumptions. 

i) {\em Type-I seesaw} extension of the SM discussed above. 
In the flavour basis, where both charged lepton and Majorana mass matrices are diagonal, 
$h$, that is in general complex, encodes all source of $C\!P$ violation
that can translate into a macroscopic $B-L$ asymmetry, injected in the form of a lepton asymmetry, 
thanks to the out-of-equilibrium decays of the very heavy RH neutrinos. A RH neutrino can decay either into
lepton and higgs doublets   with rate $\Gamma_i$ or  into anti-leptons and (h.c.) higgs doublets with rate
$\bar{\G}_i$. The two rates are in general because of $C\!P$ violation and one can define
the total $C\!P$ asymmetries
\be
\ve_i \equiv - {\Gamma_i - \bar{\Gamma}_i \over \bar{\Gamma}_i + \bar{\Gamma}_i} \,  .
\ee
Each decay of a RH neutrino $N_i$ will then generate on average a $B-L$ asymmetry, in the form 
of lepton asymmetry, given by $\ve_i$. However, the final $B-L$ asymmetry has to take into account
also the inverse processes that wash-out the asymmetry produced from decays and moreover this wash-out
can also be flavour dependent. In general we can anyway write that the final $B-L$ asymmetry will be the sum of the contributions from each RH neutrino species, so that $N_{B-L}^{\rm f} = \sum_i \, N^{(i){\rm f}}_{B-L}$.

ii) {\em Thermal production} of the RH neutrinos
in the early Universe. This implies a reheat temperature at the end of inflation $T_{RH} \gtrsim 
T_{\rm lep} = M_i/z_B$, where $M_i$ is the mass of the RH neutrino whose decays dominantly produce the
asymmetry and $z_B =2\mbox{---}10$ is a factor taking into account that the surviving asymmetry is 
generated in a relatively sharp range of temperatures below $M_i$, about the value corresponding to a departure
of equilibrium, while at higher temperature all asymmetry is quite efficiently washed-out. This is true for the
a production in a (mildly) strong wash-out regime that however is strongly favoured (and desirable) by the
measured values of the solar and atmospheric neutrino mass scales. 

A necessary condition for successful leptogenesis is $T_{\rm lep} \gtrsim T_{\rm sph}^{\rm off}$,
where $T_{\rm sph}^{\rm off} \simeq 140\,{\rm GeV}$ 
is the temperature below which sphaleron processes switch off and
go out-of-equilibrium (i.e. when $\G_{\rm sph} \lesssim H$ where $H$ is the expansion rate)
\cite{krs}.  In this way sphalerons can convert part of the lepton asymmetry into a baryon asymmetry
conserving the $B-L$ asymmetry and one obtains for the baryon-to-photon number
ratio predicted by leptogenesis
\be
\eta_{B,0}^{\rm lep} = a_{\rm sph}\,{N^{\rm f}_{B-L} \over N_{\g,0}} \,   ,
\ee
where $a_{\rm sph} \simeq 1/3$ is the fraction of 
$B-L$ asymmetry that ends up into a baryon asymmetry. 
Successful leptogenesis of course requires $\eta_{B,0}^{\rm lep} = \eta_{B,0}^{(CMB)}$.

\subsection{A problem with too many parameters}

The seesaw parameter space contains 18 additional parameters: 3 RH neutrino masses and 15 additional
parameters in the Dirac mass matrix.  Thanks to the seesaw formula, the 15 parameters in the Dirac mass
matrix can be re-expressed through the 9 low energy neutrino parameters, 3 light neutrino masses
and 6 parameters in $U$, the 3 $M_i$ and 6 parameters in a orthogonal matrix $\O$, explicitly
\be
m_D = U\,D_m^{1/2}\,\Omega\,D_M^{1/2} \,  .
\ee
The orthogonal matrix $\O$ \cite{casasibarra} encodes information on the 3 lifetimes and the 
3 total $C\!P$ asymmetries of the RH neutrinos.  Therefore, low energy neutrino experiments
by themselves cannot test the seesaw mechanism. The baryon-to-photon number ratio calculated
from leptogenesis, $\eta_B^{\rm lep}$, depends on all 18 seesaw parameters in general. The 
successful leptogenesis condition is conceptually very important since introduces a constraint on 
the RH neutrino parameters: with leptogenesis we are able to read the result of a very special
experiment occurred in the early Universe, the origin of matter, getting information on the physics at those very high energies. Model independently, leptogenesis is  insufficient to over-constraint the seesaw parameter space providing a conclusive phenomenological test but a few things might help reducing the number of independent parameters:
\begin{itemize}
\item Successful leptogenesis might be satisfied only about {\em peaks}, i.e. only for very special regions in parameter space
that can correspond to testable constraints on some  low energy neutrino parameter;
\item Some of the parameters might cancel out in the calculation of $\eta_B^{\rm lep}$;
\item Imposing some cosmologically motivated condition  to be respected such as the {\em strong thermal leptogenesis} (independence of the initial conditions) or, even stronger, that one of the heavy RH neutrino 
species is the dark matter candidate;
\item Adding particle physics phenomenological constraints, 
such as collider signatures, charged LFV, EDM's, \dots;
\item Embedding the seesaw within a model leading to conditions on $m_D$ and $M_i$. 
\end{itemize}

\subsection{Vanilla leptogenesis}

A particular successful scenario that well illustrates the possible above mentioned strategies to reduce the number of parameters  in order to obtain testable constraints or predictions on observables, is represented by 
so called {\em vanilla leptogenesis}.  It relies on the following set of assumptions: 
\begin{itemize}
\item[i)] The flavour composition of the final leptons does not influence the calculation of the final asymmetry; 
\item[ii)] Hierarchical RH neutrino spectrum ($M_2 \gtrsim M_1$); 
\item[iii)] The asymmetry produced by the heavier RH neutrino is negligible. 
\item[iv)] A set of Boltzmann (momentum integrated) rate equations  fairly describes the kinetic evolution. 
\end{itemize}
Under these four assumptions the predicted baryon-to-photon number ratio
gets a contribution only by the lightest RH neutrino decays and one obtains a very
simple expression \cite{pedestrians}, 
\be\label{exponential}
\eta_B^{\rm lep} \simeq 0.01\,\varepsilon_1 \, \kappa^{\rm f}(K_1)\,
\exp\left[-{\omega \over z_B}\,{M_1 \over 10^{10}\,{\rm GeV}}\,
{\sum_i m_i^2 \over {\rm eV}^2} \right] \,   ,
\ee
where the final {\em efficiency factor} $\k^{\rm f}(K_1)$ depends only on the lightest
RH neutrino decay parameter $K_1 \equiv \widetilde{\Gamma}_1/H(T=M_1)$, with
$\widetilde{\Gamma}_1$ indicating the lightest RH neutrino decay width. It is basically
corresponding to the number of RH neutrinos decaying out-of-equilibrium. 
The exponential factor is an effect of $\D L =2$ wash-out processes and $\omega \simeq 0.186$
while $z_B\simeq 2$ ---$10$ has a  logarithmic dependence on $K_1$. 
If in addition to the four above mentioned assumptions one 
also  v) bars fine-tuned cancellations in the see-saw formula, one obtains 
the upper bound \cite{di}
\be
\ve_1 \lesssim 10^{-6}\,{M_1 \over 10^{10}\,{\rm GeV}}\,{m_{\rm atm} \over m_1 + m_3} \,  .
\ee 
When these results are combined, from the successful leptogenesis condition one finds
a lower bound $M_1 \gtrsim 10^9\,{\rm GeV}$ \cite{di,cmb} and an upper bound 
$m_1 \lesssim 0.1 \,{\rm eV}$ \cite{upperbound,pedestrians}
that is mainly the consequence of the $\D L =2$ wash-out exponential suppression in Eq.~(\ref{exponential})
and is now interestingly confirmed by the above mentioned cosmological upper bound, 
$m_1 \lesssim 70\,{\rm meV}$,  placed by the {\em Planck} collaboration.
This upper bound is also  very interesting, since it provides an example of how, despite one starts from 18 parameters,  the successful leptogenesis condition can indeed produce testable constraints. The reason
is that the final asymmetry in vanilla leptogenesis does not depend on the 6 parameters in $U$, since this 
cancels out in $\ve_1$, and on the 6 parameters associated to the two heavier RH neutrinos. There are
only 6 parameters left ($m_1, m_{\rm atm}, m_{\rm sol}, M_1, \O^2_{11}$) out of which two are measured
thus leaving only 4 free parameters. The asymmetry however has a peak strongly suppressed by 
the value of $m_1$, due mainly to the exponential suppression from $\D L=2$
wash-out processes in Eq.~(\ref{exponential}),  that is where the upper bound on $m_1$ comes from. 

Another interesting feature of {\em vanilla leptogenesis} is that the value of the decay parameter $K_1$
varies typically within a range $\sim 10$--$50$, where the wash-out is moderately strong: not too strong to prevent
successful leptogenesis but strong enough to wash-out a large pre-existing asymmetry (including
an asymmetry generated by the heavier RH neutrinos) since its relic value is given by
\be
N_{B-L}^{\rm pre-ex,f} = N_{B-L}^{\rm pre-ex,i}\,{\rm exp}\left[-{3\pi\over 8}\,K_1 \right] \,  .
\ee
There is, however, a special region in the parameter space for which $K_1 \lesssim 1$
and in this case the observed asymmetry can be reproduced by the contribution from $N_2$ decays
realising a {\em $N_2$-dominated scenario of leptogenesis} \cite{geometry}.  Within an unflavoured description, this has to be regarded as an exception to the vanilla leptogenesis scenario that is strictly $N_1$-dominated. 
We will see, however, how accounting for flavour effects 
this scenario becomes much more important and easy to realise. 

An unpleasant feature of vanilla leptogenesis is that, imposing so called 
$SO(10)$-inspired conditions with $V_L \simeq V_{CKM}$,  and 
$(m_{D1}, m_{D2}, m_{D3}) \sim (m_{\rm up}, m_{\rm charm}, m_{\rm top})$,
and barring very fine-tuned crossing level solutions,
one has $M_1 \sim 10^{5}\,{\rm GeV}$, well below the lower bound on $M_1$ 
for successful $N_1$-dominated leptogenesis. Moreover also the $N_2$-dominated leptogenesis
scenario cannot be realised since in $SO(10)$-inspired leptogenesis one has strictly $K_1 \gg 1$. 
We will be back  on $SO(10)$-inspired leptogenesis and see how this problem can be circumvented. 

\section{Flavour effects}

The existence of a stringent lower bound on $M_1$ has been one of the main motivations to investigate
scenarios of leptogenesis beyond vanilla leptogenesis in the last years. There have been four main directions:
\begin{itemize}
\item Leptogenesis with quasi-degenerate RH neutrino spectrum leading to a resonant enhancement
of the $C\!P$ asymmetry (resonant leptogenesis) \cite{resonant}.
\item Beyond the minimal scenario of leptogenesis, either considering non-minimal versions of the 
seesaw mechanism (such as type II see-saw, inverse see-saw, double seesaw) \cite{hambye} or relaxing the thermal RH neutrino production  assumption considering non-thermal leptogenesis scenarios \cite{nonthermal}.
\item Improved kinetic description beyond simple rate Boltzmann equation: momentum
dependence \cite{momentum}, density matrix equations \cite{bcst,zeno}, 
Kadanoff-Baym and closely related closed-time path formalism \cite{qke}.
\item Charged lepton and heavy neutrino flavour effects and their interplay. 
\end{itemize} 
All these four directions have stimulated intense investigations providing a much deeper insight 
into the calculation of the asymmetry in leptogenesis and the possibility to evade the
bounds from the vanilla leptogenesis scenario. However, the most far-reaching implications, 
certainly in connection
with models and low energy neutrino experiments, are those from flavour effects
and for this reason here we focus on this particularly important development. 

\subsection{Charged lepton flavour effects}

Let us first consider the $N_1$-dominated scenario. 
If $5 \times 10^8\,{\rm GeV} \lesssim M_1 \lesssim 5 \times 10^{11}\,{\rm GeV}$
then the flavour composition of the leptons (and anti-leptons) produced in the $N_1$-decays influence
the value of the final asymmetry since leptons have to be described as an incoherent mixture of a 
tauon component and, of a coherent superposition of the electron and muon components due to the 
fast $\tau$-interactions \cite{bcst,flavour}.
 In this situation a {\em two-flavoured regime} is realised and the final asymmetry
has to be calculated as the sum of a tauon component and of a electron+muon component, since the two in general
experience a difference wash-out, i.e. a different kinetic evolution. An approximated expression valid for
$K_1 \gg 1$ in this case is given by \cite{predictions}
\be
N_{B-L}^{\rm f} \simeq 2\,\ve_1 \, \k(K_1) + {\D p_{1\tau} \over 2}
\,\left[\kappa(K_{1 e+\m}) - \k(K_{1\t}) \right] \,  ,
\ee
where $K_{1\a}$ ($\a=e,\m,\t$) are the decay flavoured parameters 
($K_{1 e+\m} \equiv K_{1e}+K_{1\m}$) and $\D p_{1\t}$ is the difference
between the probability that a lepton produced by $N_1$ decays 
is in a tauon flavour and the probability that the anti-lepton produced by $N_1$-decays
is in a anti-tauon flavour: it is a measurement of $C\!P$ flavour violation. If the second term vanishes,
then the inclusion of flavour effects simply double the asymmetry but if the second term is non-vanishing then
there can be much more significant implications. In particular now even though $\ve_1 =0$
one can still produce the correct asymmetry and in this respect low energy phases 
are crucial. More generally, the second term is now strongly 
depending on low energy neutrino parameters. In particular Majorana phases play quite an important
role in establishing whether  the difference between the two efficiency factors cancel out, so that 
the second term is suppressed, or whether one dominates over the other so that there is no suppression
\cite{predictions}.
In this way the term $ \propto \D p_{1\t}$ introduces an additional (flavoured) source of $C\!P$ violation 
that in some cases can be dominant.  

If $M_1 \lesssim 5\times 10^8\,{\rm GeV}$ then also muon interactions
are fast enough to break the coherence of the electron-muon component and 
one has to consider a three flavoured regime where the final asymmetry has to be calculated
as the sum of three different contributions from each charged lepton flavour.  However,
in a $N_1$-dominated scenario with hierarchical spectrum for these $M_1$ values one cannot reproduce
the observed asymmetry and for this reason the three-flavour regime is less relevant. 
\footnote{This conclusion holds in a non-supersymmetric framework. In a supersymmetric framework
the transition between a two-flavoured and a three-flavoured regime occurs at values 
$M_1 \simeq 5\times 10^8\,{\rm GeV}\,(1+\tan^2 \beta)$ and for $\tan^2 \b$ large enough successful leptogenesis can be attained even in a three-flavoured regime.}

\subsection{Heavy neutrino flavour effects}

In general one should also consider the asymmetry produced by the 
out-of-equilibrium decays of the heavier RH neutrinos.
In an unflavoured approximation, one would obtain that this is efficiently 
washed-out and can be neglected except for a special region in parameter space \cite{geometry}.
However, when charged lepton flavour effects are considered, the wash-out has to be considered
along different flavour directions and is in general reduced \cite{vives}. Even when all three masses
are above $10^{12}\,{\rm GeV}$ and charged lepton effects are absent, one still has to consider
that a lighter RH neutrino $N_i$ can only wash-out the asymmetry along the ${\ell}_i$
flavour direction but not along the orthogonal direction in flavour space \cite{bcst,engelhard}.

When both charged lepton and heavy neutrino flavour effects are considered, one has 
to consider 10 different RH neutrino hierarchical 
mass spectra, shown in Fig.~1, giving rise to different expressions for the calculation of the final asymmetry with
Boltzmann equations \cite{dm2bol}. 
\begin{figure}
\begin{minipage}{140mm}
\begin{center}
\includegraphics[width=24mm]{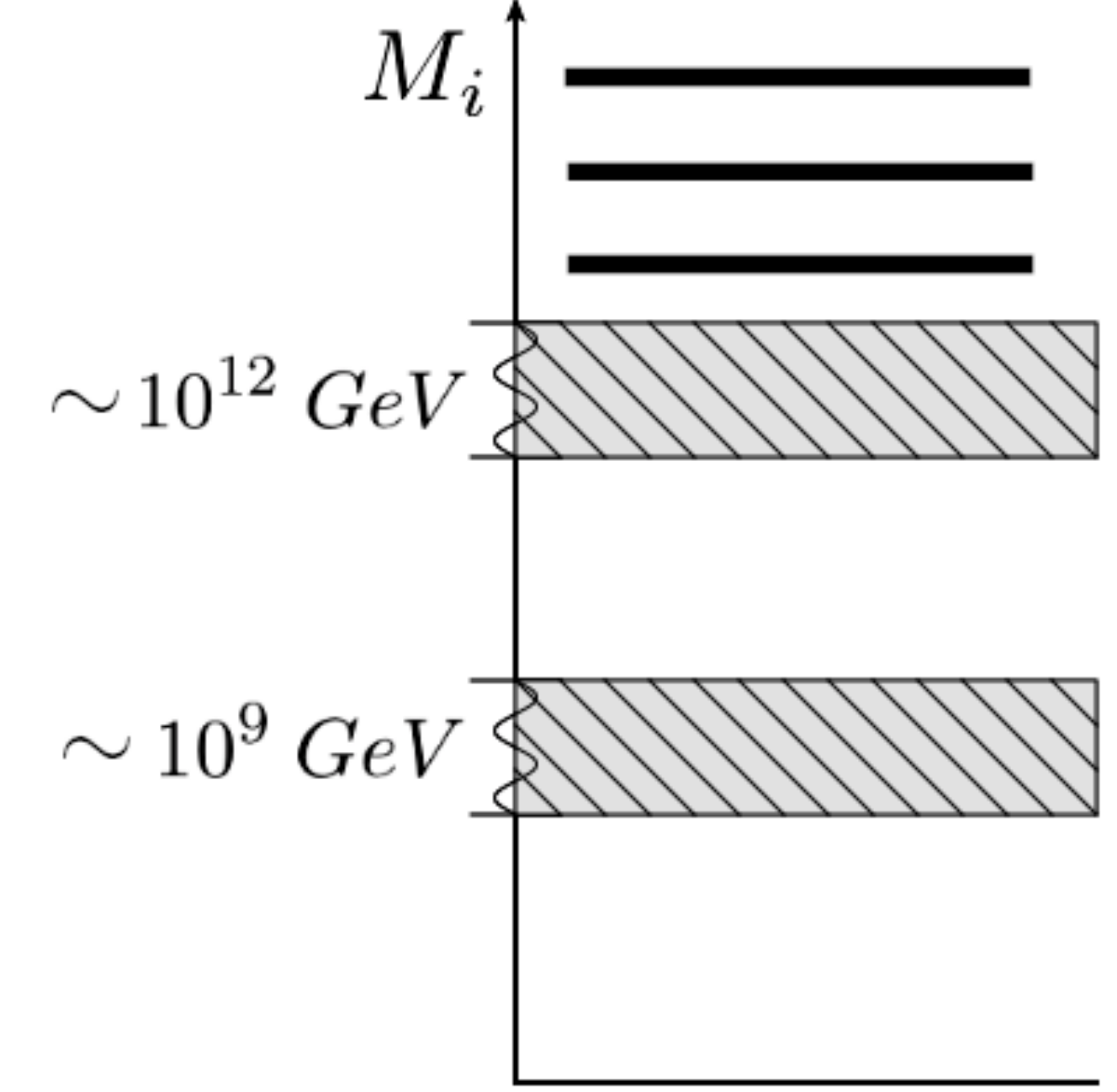} \hspace*{2mm}
\includegraphics[width=24mm]{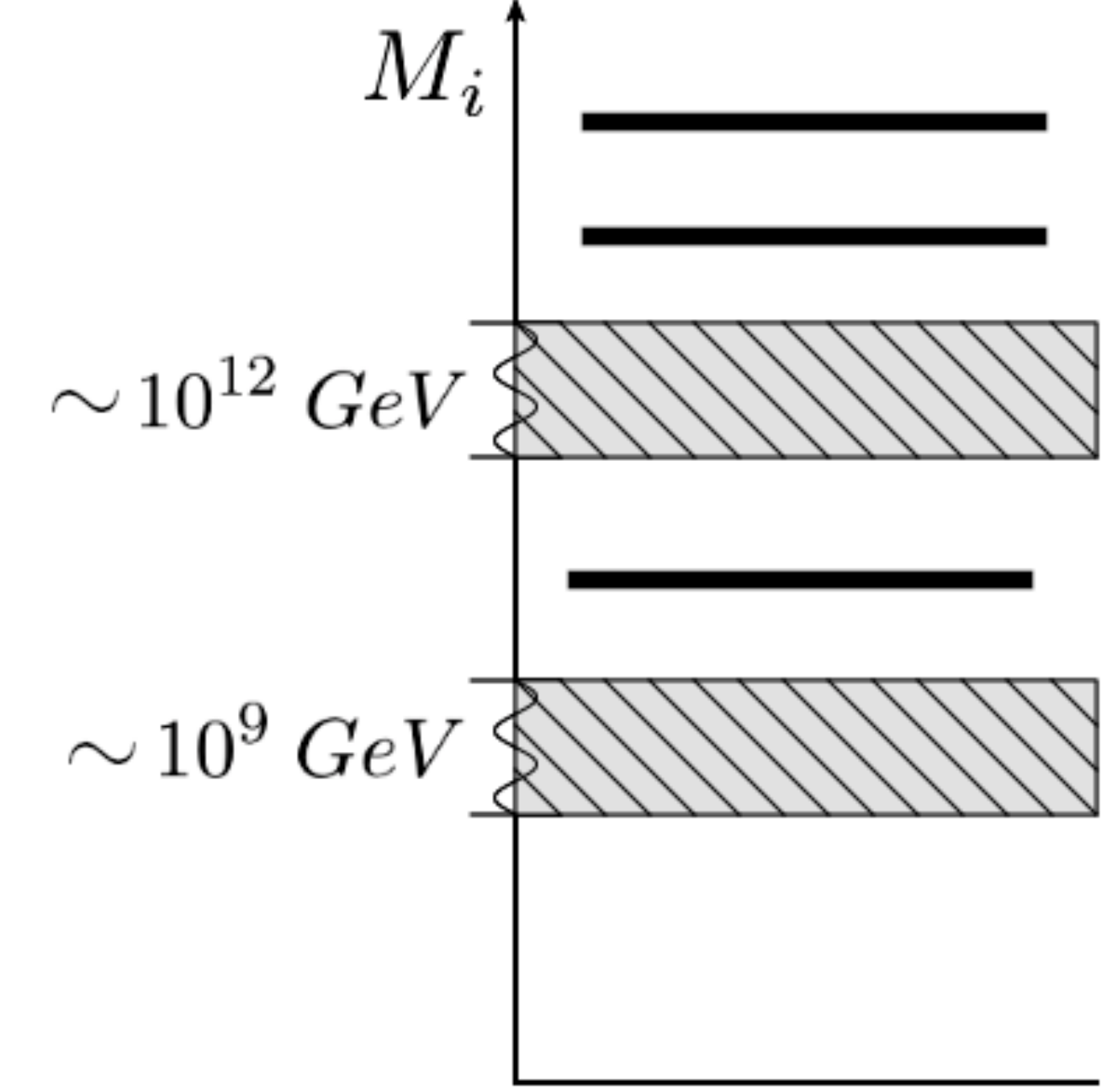} \hspace*{2mm}
\includegraphics[width=14mm]{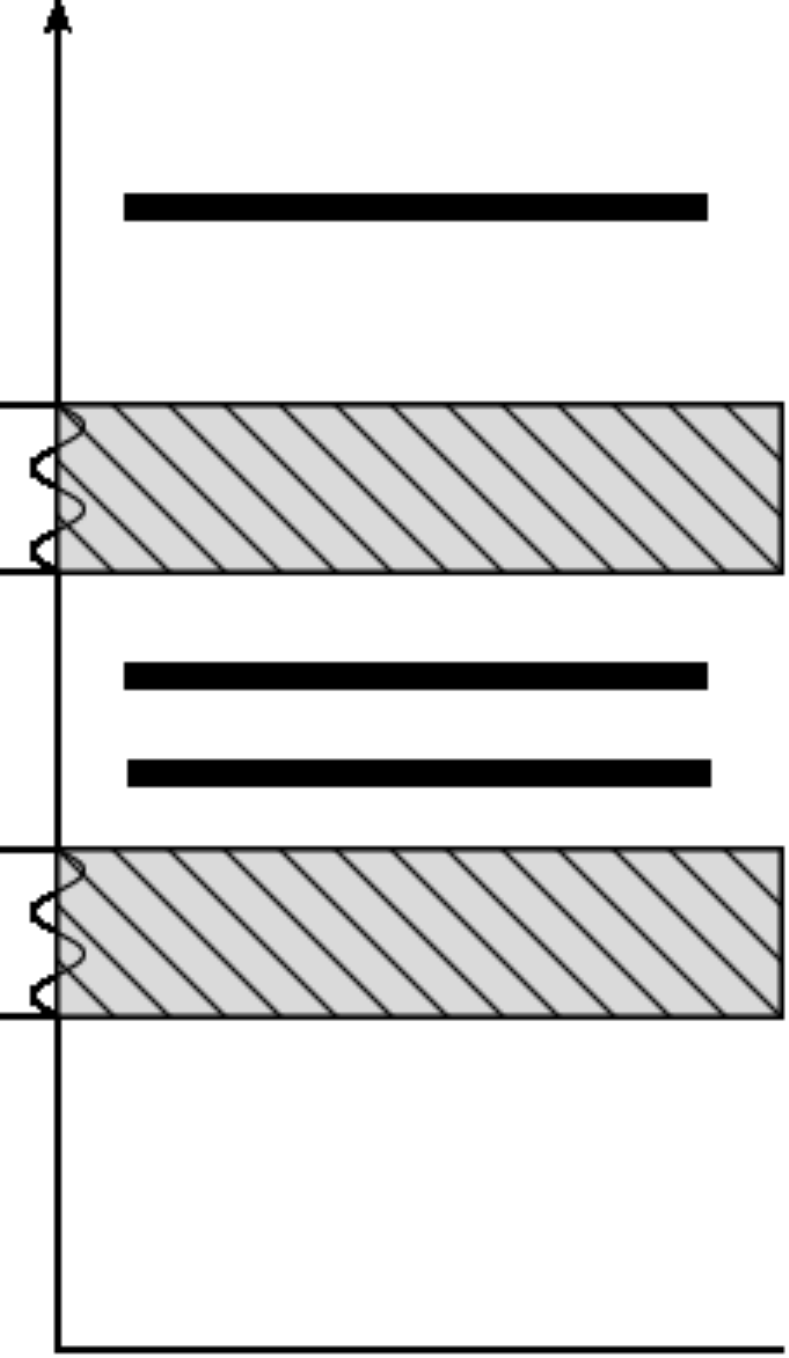} \hspace*{2mm}
\includegraphics[width=14mm]{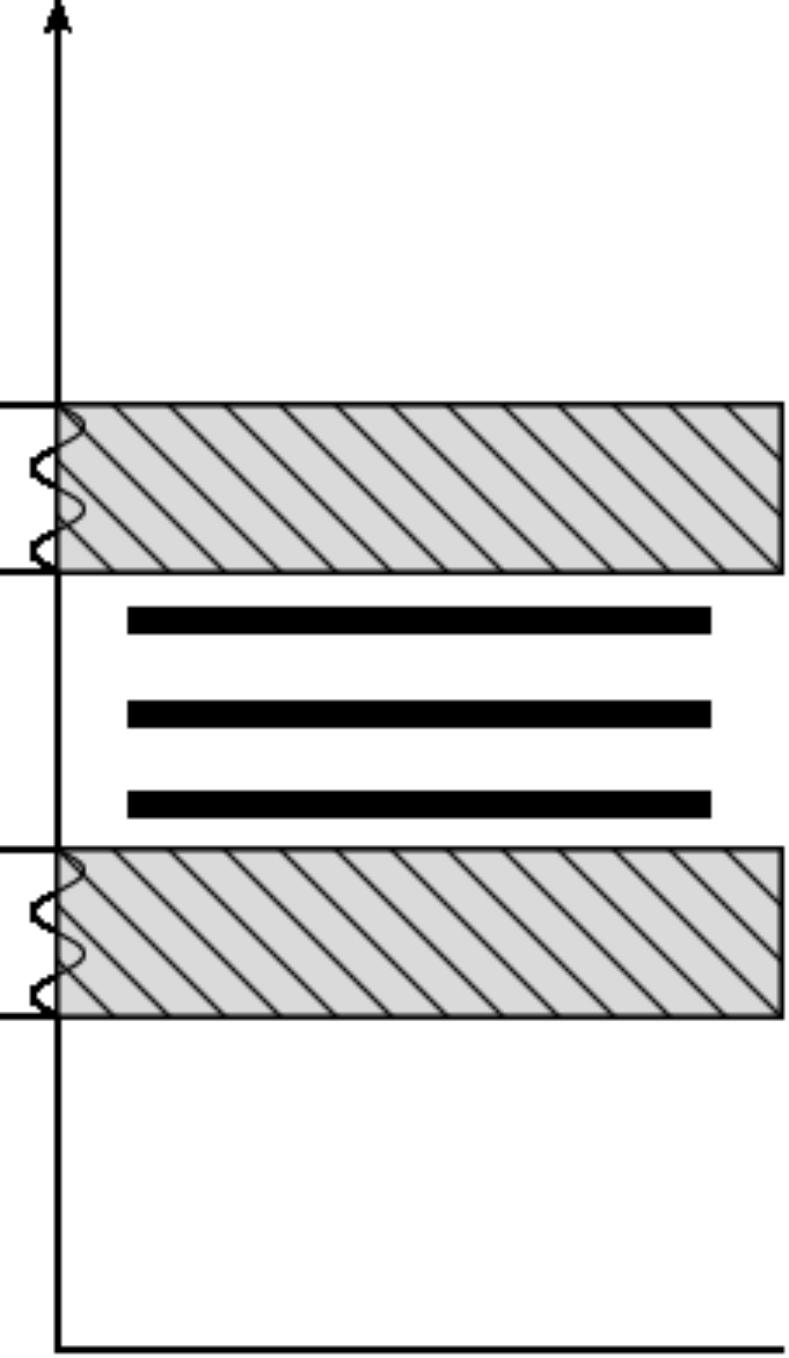}
\end{center}
\begin{center}
\includegraphics[width=24mm]{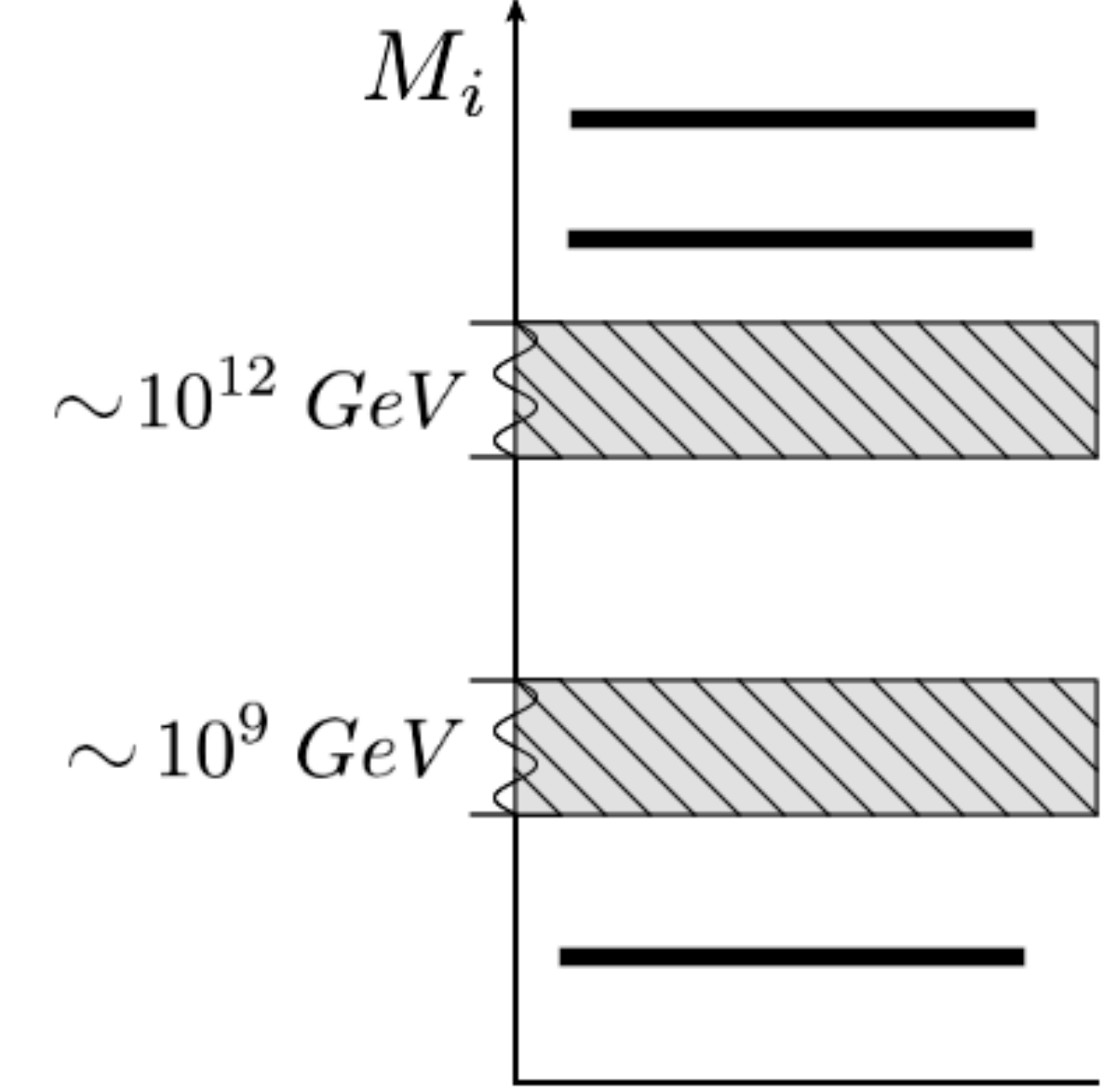} \hspace*{2mm}
\includegraphics[width=14mm]{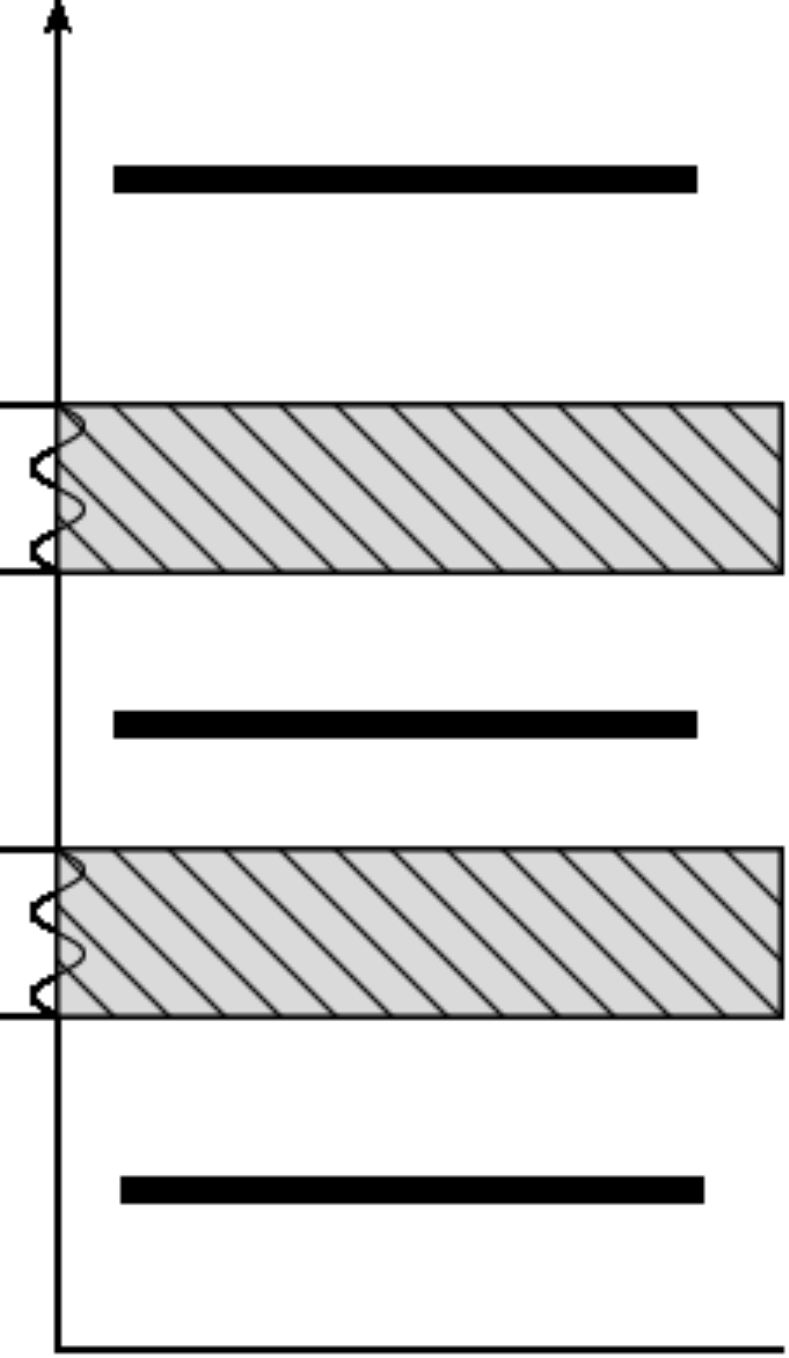} \hspace*{2mm}
\includegraphics[width=14mm]{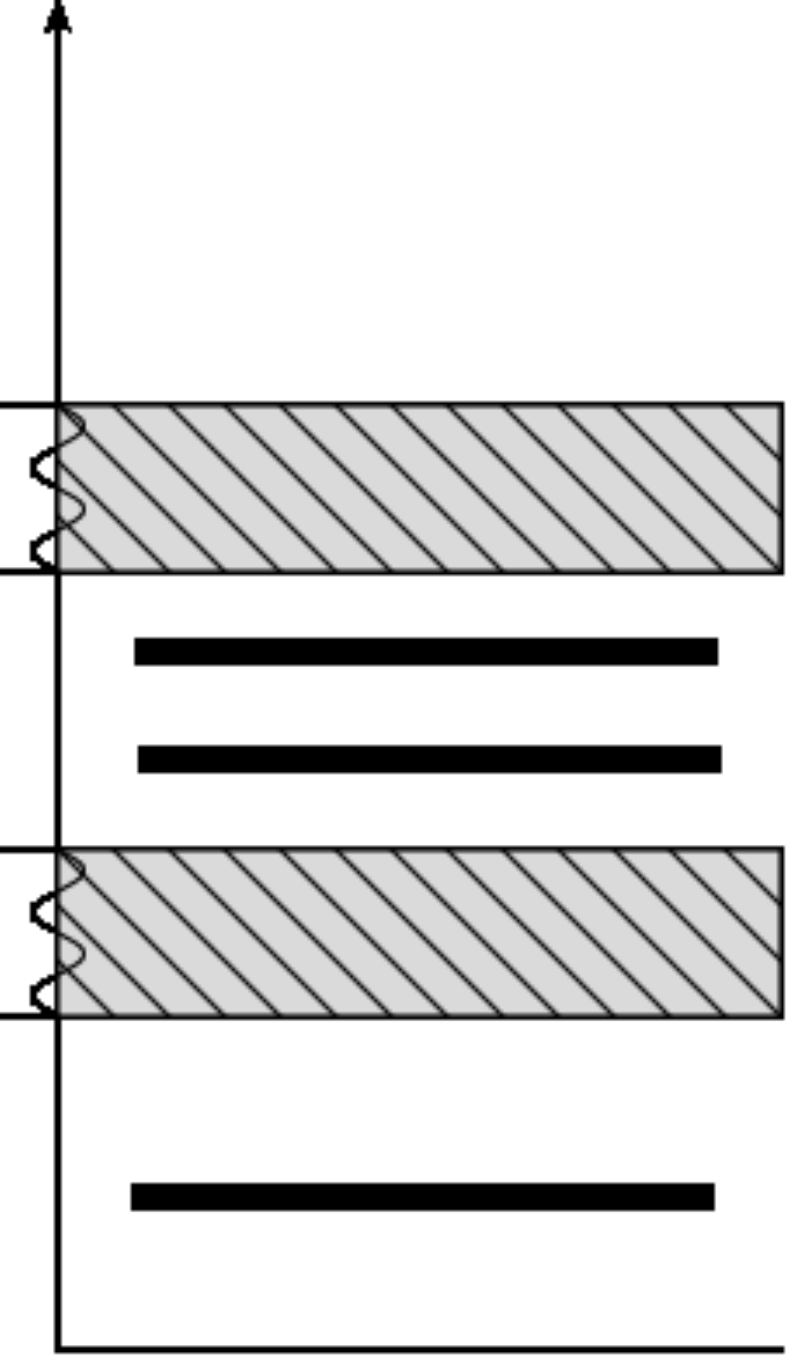} \hspace*{2mm}
\end{center}
\begin{center}
\includegraphics[width=24mm]{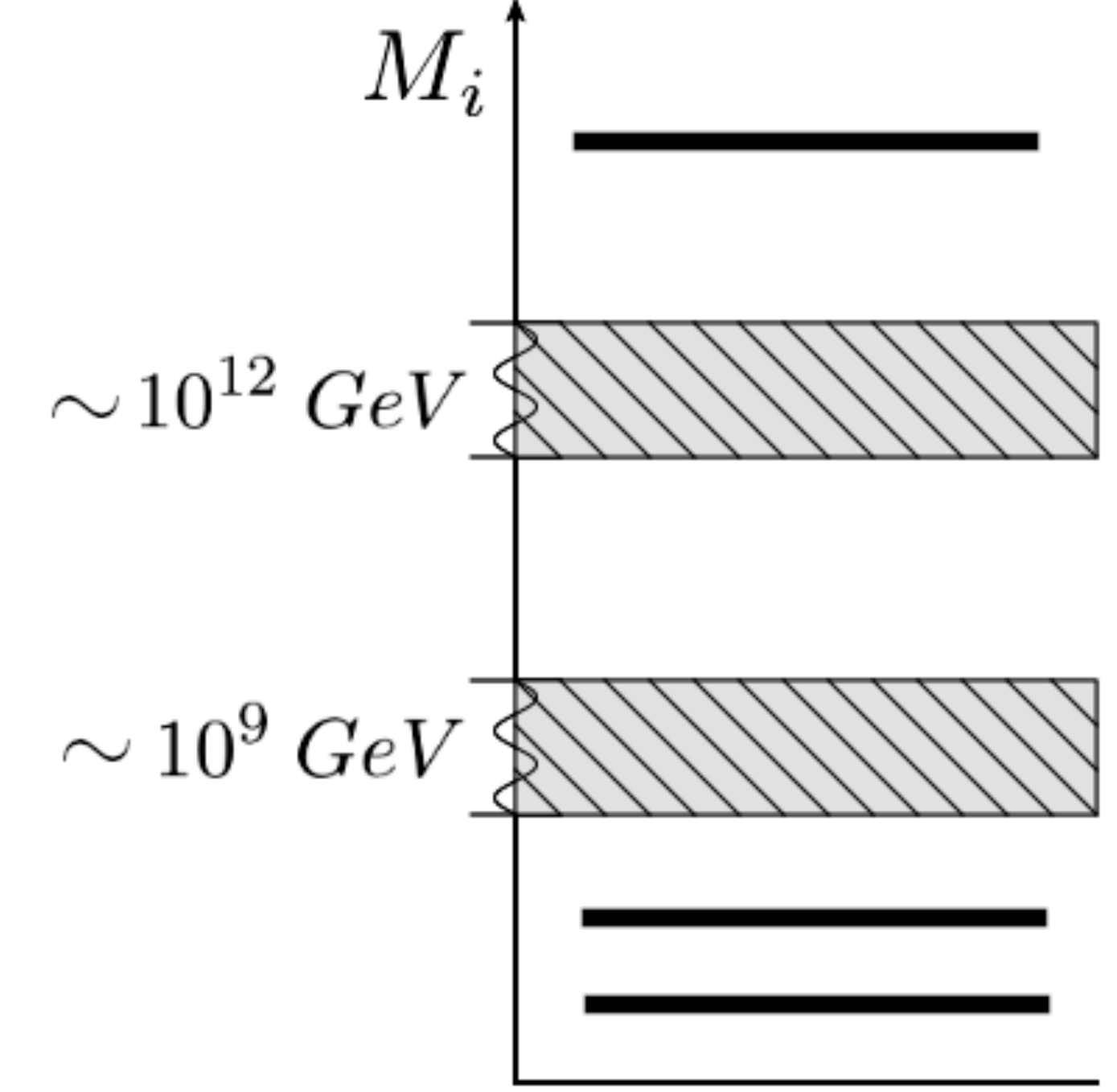} \hspace*{2mm}
\includegraphics[width=14mm]{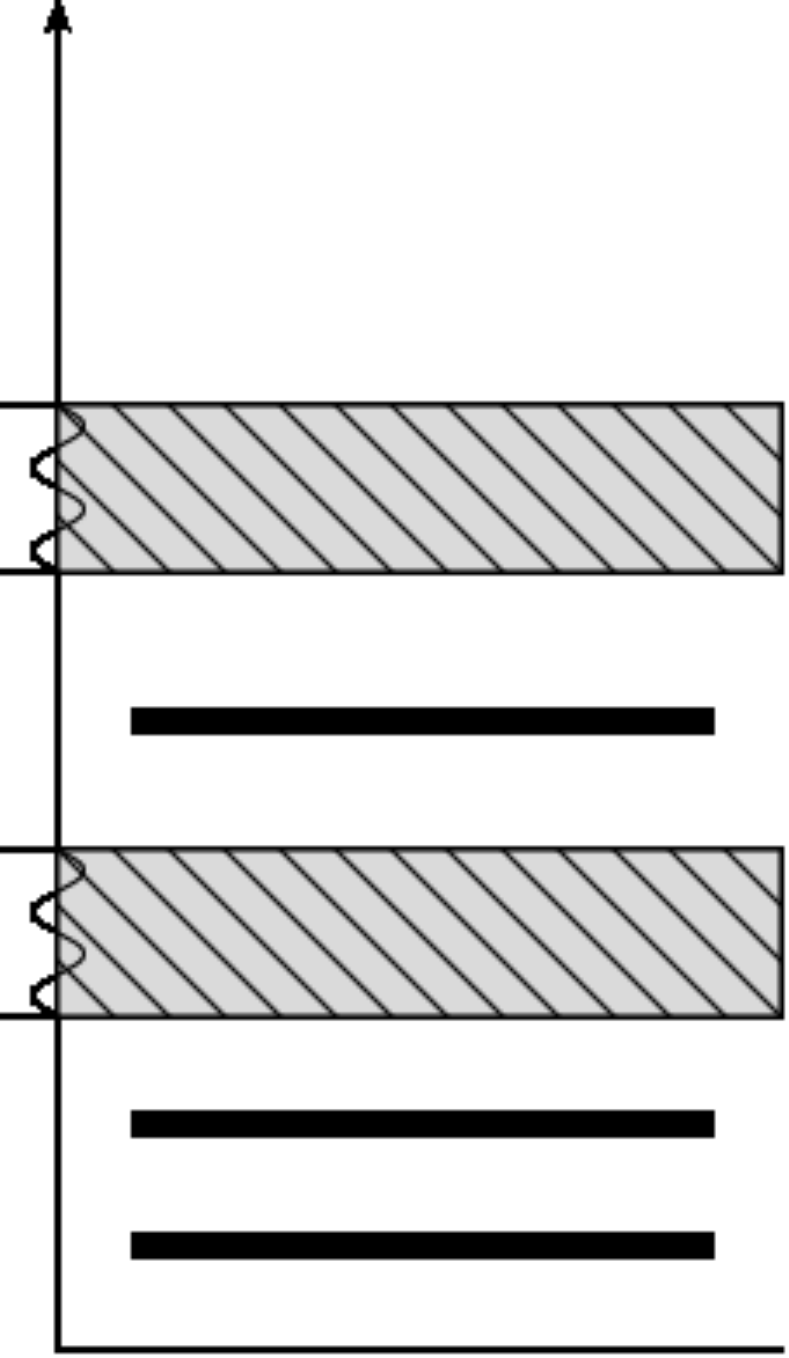} \hspace*{2mm}
\includegraphics[width=14mm]{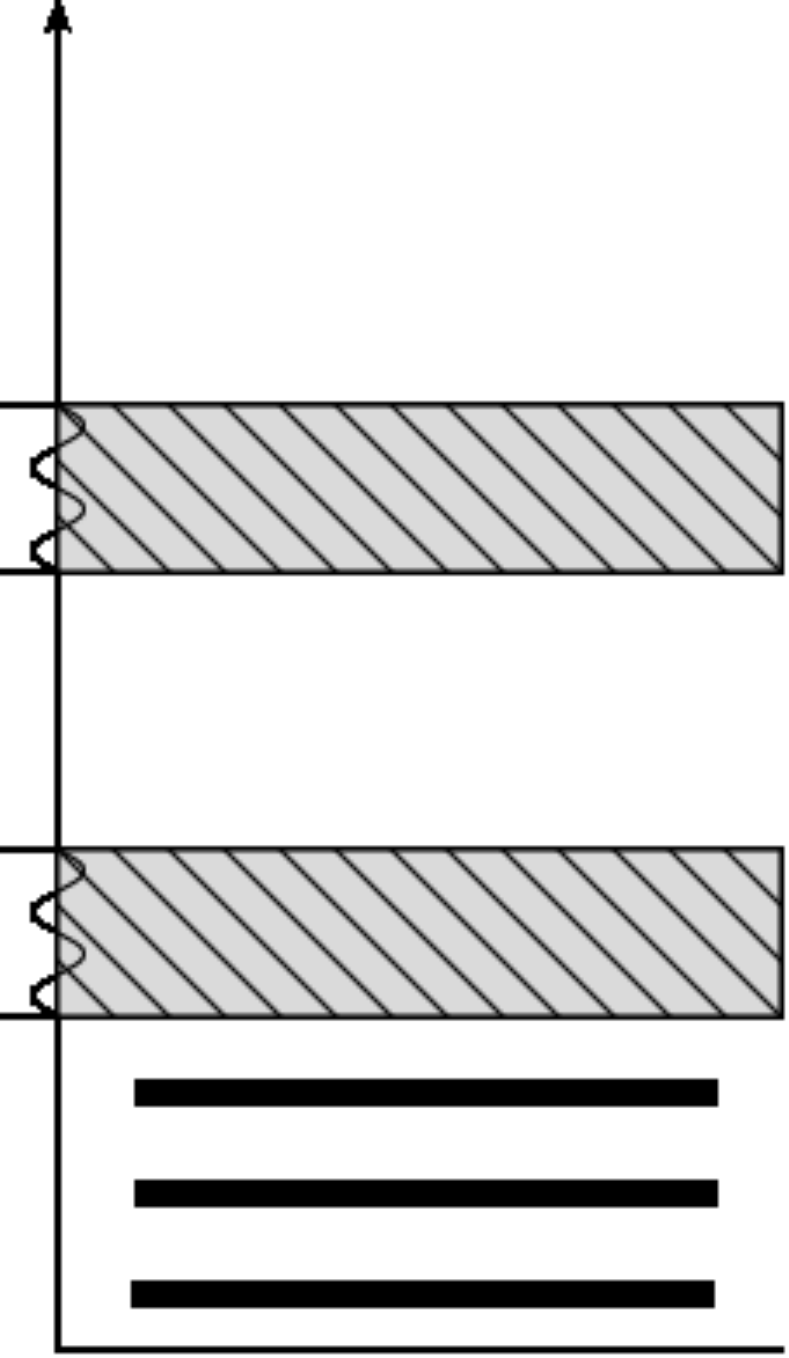}
\end{center}
\caption{The 10 RH neutrino mass patterns corresponding to
leptogenesis scenarios with
different sets of classical Boltzmann equations for the
calculation of the final asymmetry.}
\end{minipage}          
\end{figure}
The dashed intervals  indicate RH neutrino mass ranges corresponding to
transition regimes  between two different fully flavoured regimes. In these 
transition regimes density matrix equations should be used instead of Boltzmann equations
applyiable in fully flavoured regimes.
The top-left panel corresponds to the heavy neutrino flavoured scenario with all three $M_i \gg 10^{12}\,{\rm GeV}$. This scenario of leptogenesis 
typically emerges within models with discrete flavour symmetries \cite{heavyflavoured}.

\subsection{$N_2$-dominated scenario}

An important case is obtained for $M_1 \ll 10^9\,{\rm GeV}$ since in this case necessarily the asymmetry has to be  generated by the two heavier RH neutrinos and typically  the one generated by the heaviest is negligible so that one obtains a $N_2$-dominated scenario (corresponding to the
three panels in the third row of Fig.~1). 
While in the unflavoured approximation it is realised only for quite a special 
choice of parameters \cite{geometry}, when flavour effects are taken into account
the region of applicability greatly enlarges \cite{vives,bounds}. 
This scenario 
(specifically the central panel in the third row) 
has two interesting features: it emerges naturally 
when $SO(10)$-inspired conditions are imposed \cite{SO10inspired}
and it is the only one that can realise independence 
of the initial conditions, quite an interesting combination of 
completely independent features \cite{problem}.

\section{$SO(10)$-inspired leptogenesis}

In the unflavoured case we have seen that imposing $SO(10)$-inspired conditions and barring fine-tuned
crossing level solutions, successful leptogenesis cannot be attained since $M_1 \ll 10^9 \,{\rm GeV}$
and at the same time an $N_2$ contribution is efficiently washed-out.
However, when flavour effects are considered, then the  $N_2$ asymmetry can escape
the $N_1$-washout for a set solutions satisfying successful leptogenesis. 
Typically the final asymmetry is in the tauon flavour \cite{SO10lep}.
Interestingly this set of solutions requires certain constraints on the low energy neutrino 
parameters. For example the lightest neutrino mass cannot be below $\simeq 1\,{\rm meV}$, i.e. one expects
some deviation form the hierarchical limit though right now we do not know any experimental way
 to fully test 
this lower bound. It should be added that $SO(10)$-inspired leptogenesis also strongly favours
normally ordered neutrino masses  and that for $m_1 \simeq m_{\rm sol} \simeq 10\,{\rm meV}$
it is allowed only for $\theta_{23}$ in the first octant. 

\subsection{Decrypting $SO(10)$-inspired leptogenesis}

Imposing $SO(10)$-inspired conditions and barring crossing level solutions, it is possible to find
quite accurate expressions for all important quantities necessary to calculate the asymmetry. We refer
the reader to \cite{SO10decryption} for a detailed discussion, here we just give the results for the 
RH neutrino masses, given by
\be
M_1    \simeq   \a_1^2 \, {m^2_{\rm up} \over |(\widetilde{m}_\nu)_{11}|} \, , \;\;
M_2  \simeq    \a_2^2 \, {m^2_{\rm charm} \over m_1 \, m_2 \, m_3 } \, {|(\widetilde{m}_{\nu})_{11}| \over |(\widetilde{m}_{\nu}^{-1})_{33}|  } \,  ,  \;\;
M_3  \simeq   \a_3^2\, {m^2_{\rm top}}\,|(\widetilde{m}_{\nu}^{-1})_{33}|  ,
\ee
where we defined $(\a_1,\a_2,\a_3) \equiv (m_{D1}/m_{\rm up}, m_{D2}/m_{\rm charm}, m_{D3}/m_{\rm top})$
and $\widetilde{m}_{\nu} \equiv V_L\,m_{\nu}\,V_L^T$.
In this way one arrives to a full analytical expression $\eta_B^{\rm lep}(m_{\nu};\a_i,V_L)$.

\subsection{Strong thermal $SO(10)$-inspired leptogenesis}

When flavour effects are taken into account there is only one scenario of (successful) leptogenesis allowing 
for independence of the initial conditions: the tauon $N_2$-dominated scenario, where the asymmetry is produced by the $N_2$ decays in the tauon flavour \cite{problem}. The conditions are quite special since it is required that a large pre-existing asymmetry is washed-out by the lightest RH neutrino in the electron and muon flavours. 
The next-to-lightest RH neutrinos both wash-out a large pre-existing tauon asymmetry  and also produce  the observed asymmetry in the same tauon flavour escaping the lightest RH neutrino wash-out.

It is then highly non trivial that this quite special set of conditions can be realised by a subset 
of the $SO(10)$-inspired solutions satisfying successful leptogenesis \cite{strongSO10}. 
For this subset the constraints
are quite stringent and they pin down a quite well defined solution: the strong $SO(10)$-inspired solution.
This is  characterised by non-vanishing reactor mixing angle, normally ordered neutrino masses, atmospheric mixing angle in the first
octant and $\d$ in the forth quadrant  ($\sin\d <0$ and $\cos\d >0$).  In addition the lightest neutrino
mass has to be within quite a narrow range of values about $m_1 \simeq 20\,{\rm meV}$ corresponding
to a sum of neutrino masses, the quantity tested by cosmological observations, 
$\sum_i m_i \simeq 95\,{\rm meV}$, implying a deviation from the normal hierarchical limit
predicting $\sum_i m_i \simeq 60\,{\rm meV}$ detectable during next years. 
At the same time the solution also predicts a $00\b\nu$ signal with $m_{ee} \simeq 0.8\, m_1 \simeq 16\,{\rm meV}$. Some of these constraints are shown in Fig.~2. 
\begin{figure}
\begin{minipage}{140mm}
\begin{center}
\includegraphics[width=44mm]{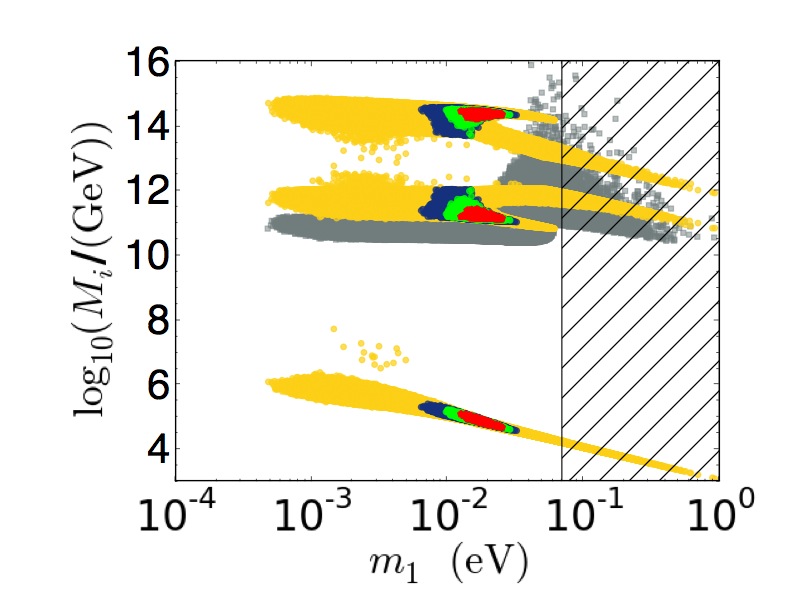} \hspace*{1mm}
\includegraphics[width=44mm]{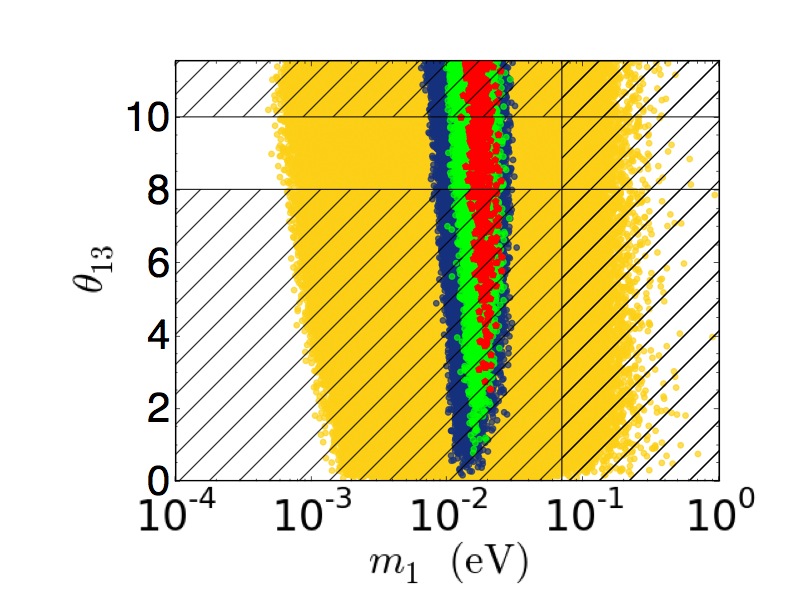} \hspace*{1mm}
\includegraphics[width=44mm]{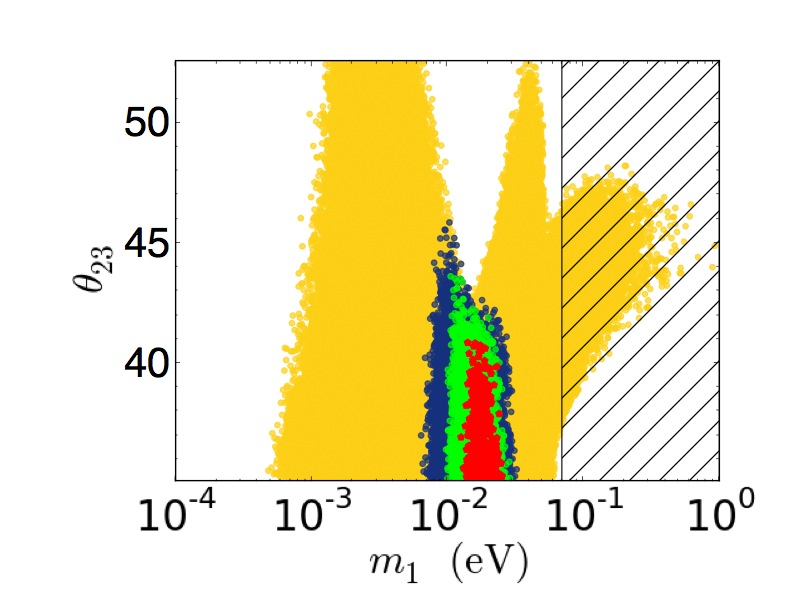} \hspace*{1mm}
\end{center}
\begin{center}
\includegraphics[width=44mm]{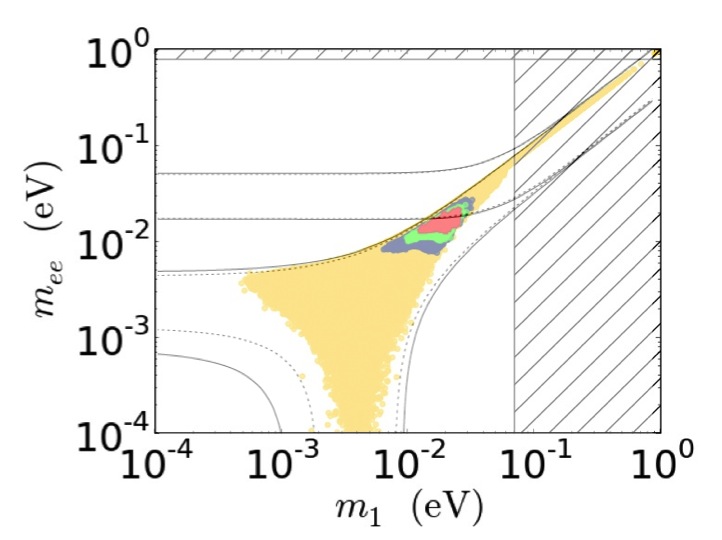} \hspace*{1mm}
\includegraphics[width=44mm]{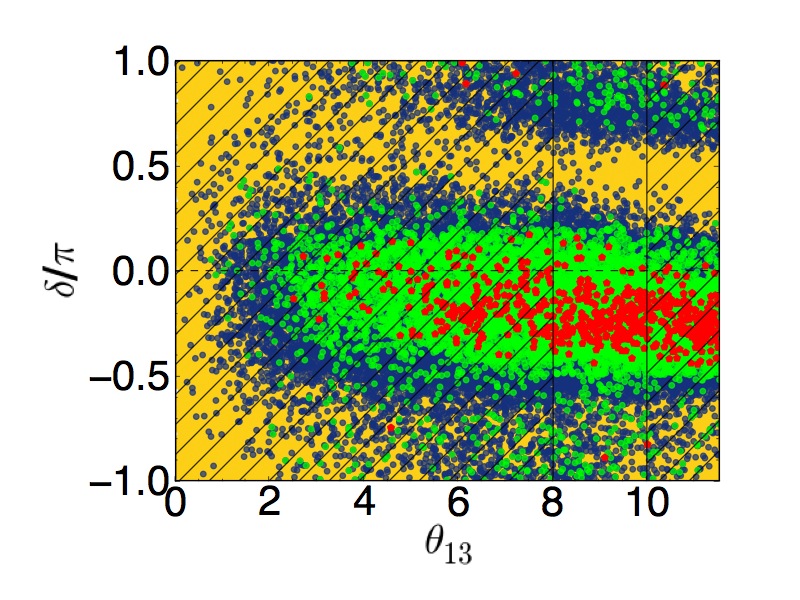} \hspace*{1mm}
\includegraphics[width=44mm]{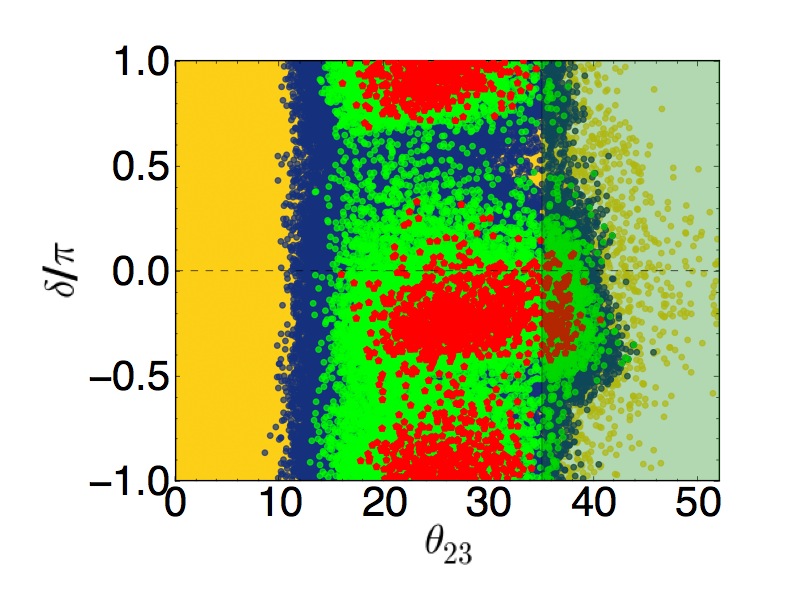}
\end{center}
\caption{Some of the constraints on low energy neutrino parameters deriving from $SO(10)$-inspired
leptogenesis (from \cite{strongSO10}). 
The yellow points are obtained imposing successful $SO(10)$-inspired leptogenesis but without 
imposing strong thermal condition (wash-out of a
large pre-existing asymmetry). The blue, green and red points correspond respectively to 
the subset of solutions also respecting the strong thermal leptogenesis (the strong $SO(10)$-inspired leptogenesis solution) respectively for 
an initial value of the pre-existing asymmetry $N_{B-L}^{\rm pre-ex,i}=10^{-3}, 10^{-2}, 10^{-1}$.}
\end{minipage}          
\end{figure}
 In light of the latest experimental results discussed in the introduction, this solution 
is quite intriguing since, in addition to rely on the same moderately strong wash-out as in vanilla
leptogenesis and due to the fact that both the solar and the atmospheric scale are $\sim 10\,{\rm meV}$,
the leptogenesis conspiracy \cite{bounds},
it has also correctly predicted  a non-vanishing reactor mixing angle and it is currently 
in very good agreement with the best fit parameters from neutrino mixing experiments (to our knowledge is
the only model that has truly predicted $\sin \d < 0$).  Notice that the possibility to have a large pre-existing
asymmetry prior to the onset of leptogenesis at the large reheat temperatures required, is quite a plausible
possibility (in particular one could have a traditional GUT baryogenesis followed by leptogenesis), so that the assumption of strong thermal leptogenesis should be regarded as a reasonable setup. 

It is also possible to consider a supersymmetric framework for $SO(10)$-inspired leptogenesis \cite{susySO10}. 
In this case the most important modification to be taken into account is that the critical values for
$M_1$ setting the transition from one flavour regime to another are enhanced by a factor $1+\tan^2\b$ \cite{flavour}
and for sufficiently large values of $\tan\b$ the production might occur in a three flavoured regime rather
than in a two-flavour regime. This typically goes in the direction of enhancing the final asymmetry since the wash-out at the production is reduced.

\subsection{Realistic models}

A first example of realistic models satisfying $SO(10)$-inspired conditions and able to fit all lepton and quark
mass and mixing parameters are of course $SO(10)$ models. A specific example is given by renormalizable $SO(10)$-models for which the Higgs fields belong to 10-, 120-, 126-dim representations yielding specific mass relations among the various fermion mass matrices \cite{SO10}.  Recently reasonable fits have been obtained typically pointing  to a compact RH neutrino spectrum with all RH neutrino masses falling in the two-flavour regime.
This compact spectrum solutions imply however huge fine-tuned cancellations in the seesaw formula.  
 Also fits realising the $N_2$-dominated scenario have been obtained \cite{rodejohann,babubajc} and in this
 case, there is no fine-tuning in the seesaw formula. 
Note  that $SO(10)$-inspired conditions can be also realised beyond $SO(10)$-models.
For example recently a Pati-Salam model combined with $A_4$ and $Z_5$ discrete symmetries has been
proposed satisfying $SO(10)$-inspired conditions \cite{A2Z}    and also successful $SO(10)$-inspired
leptogenesis \cite{A2Zlep}. On the other hand a realistic model realising strong thermal
$SO(10)$-inspired leptogenesis has not yet been found.  

\section{Two RH neutrino models and dark matter}

Another popular scenario of leptogenesis is realised within a 2 RH neutrino model \cite{2RHN}
(corresponding to the third panel in Fig.~1). In this case
the heaviest RH neutrino has a mass $M_3 \gg 10^{15}\,{\rm GeV}$ and it effectively decouples from the
seesaw formula. In this case there is a lower bound $M_1 \gtrsim 2 \times 10^{10}\,{\rm GeV}$ 
from leptogenesis \cite{2RHNlep}. Also in this case  there are regions in the parameters space, though more special, where the $N_2$-production is essential to get successful leptogenesis. 

Recently a realistic 2 RH neutrino scenario of leptogenesis has been shown to emerge within
a $A4 \times SU(5) $ SUSY  GUT model \cite{bjorkeroth1} and also within a
$\Delta(27) \times  SO(10)$ model \cite{bjorkeroth2}, showing that a 
a $SO(10)$ model combined with a discrete symmetry does not necessarily 
give rise to a very hierarchical spectrum of RH neutrino masses with
 $M_1\ll 10^9 \, {\rm GeV}$.  

Intriguingly, within a 2 RH neutrino seesaw scenario, one can also consider the case when the third RH neutrino
decouples from the seesaw formula not because it is very heavy but because its Yukawa coupling is very small.
In the limit when it basically vanishes, the RH neutrino becomes stable and can play the role of dark matter
\cite{anisimov}.  The difficulty is to find a plausible production mechanism. A minimal way the does not require to 
lower all neutrino Yukawa couplings in order to enhance the LH-RH neutrino mixing as in the $\nu$MSM model,
is to introduce a non-renormalizable Higgs portal-like operator 
$(\l_{ij}/\Lambda)\,\phi\,\phi^\dagger\,N_i\,\bar{N}_j^c$, where $\L$ is the scale of new physics (or a combination of more scales).  The very interesting feature of this operator is that it can at the same time
be responsible for the RH neutrino production through Landau-Zener non adiabatic resonant 
conversion, enhancing medium effects, and at the same time make the RH neutrino unstable. Interestingly 
an allowed region exists such that both requirements of production and stability on cosmological scales
can be satisfied and this region is for a mass
of the DM RH neutrino in the range ${\rm EeV} \gtrsim M_{DM} \gtrsim {\rm TeV}$. 
However at the same time one can have some very high energy neutrino flux  that
might give a detectable contribution at IceCube \cite{unified}. 
\footnote{Interestingly an excess at energies ${\cal O}(10\,{\rm TeV}-100\,{\rm TeV})$ with respect to an astrophysical component has been recently discussed and interpreted in terms of a decaying DM \cite{morisi}, though FermiLat $\g$-rays observations seem to constraint such a contribution. \cite{murase}}
The scenario is also
compatible with resonant leptogenesis in a two RH neutrino model, realising in this way a unified picture of 
leptogenesis and dark matter that will be tested in next years at IceCube. 

\section{Conclusions}
Despite the absence of new physics
at colliders so far, with neutrino physics and cosmology (and hopefully with 
high energy neutrinos at Neutrino Telescopes) 
there are well motivated ways in the next years to disclose the nature of the SM extension 
that is necessary in order to explain neutrino masses and the cosmological puzzles. 
High energy seesaw models embedded within GUT theories provide a very simple and attractive
way in this respect to address the matter-antimatter asymmetry and dark matter
of the Universe with testable predictions. 

\subsection{Acknowledgments}

I wish to thank Steve Blanchet, Steve King, Patrick Ludl,  Luca Marzola, 
Sergio Palomares-Ruiz, Michele Re Fiorentin, for a fruitful collaboration on the subjects discussed in this talk.  
It is also a pleasure to thank the organisers of Neutrino 2016 and NuFact2016 for the nice invitation and the opportunity to present this talk.



\end{document}